# Light Energy Driven Nanocommunications with FRET in Photosynthetic Systems

**Aleksandra Orzechowska[1], Joanna Fiedor[1], and Pawel Kulakowski[2]**
[1]AGH University of Science and Technology, Faculty of Physics and Applied Computer Science, al. Mickiewicza 30, 30-059 Krakow, Poland
[2]AGH University of Science and Technology, Institute of Telecommunications, al. Mickiewicza 30, 30-059, Krakow, Poland

Corresponding author: Pawel Kulakowski. (e-mail: kulakowski@agh.edu.pl).

This work was supported in part by the funds of Faculty of Physics and Applied Computer Science (No. 16.16.220.842) and the Faculty of Computer Science and Telecommunications, AGH University of Science and Technology, within the subventions of the Polish Ministry of Science and Higher Education.

**ABSTRACT** Chlorophylls and carotenoids, key components of photosynthetic systems, are proposed for molecular communications at the nanoscale with the mechanism of resonance energy transfer. Both types of pigments are introduced focusing on their exceptional properties like energy harvesting, ultra-low energy consumption during transmissions, picoseconds delays, an ability for signal conversions, and bio-compatibility. The theoretical considerations are supported by two spectroscopic experiments on the photosystem II complex. The first experiment aims at the description of the photosystem II including calculation of the energy transfer efficiency between carotenoids and chlorophylls. In the second one, the photochemical efficiency is estimated, showing how effective the chlorophylls are in further energy processing. With the experimentally determined values, communication and energetic performance are analyzed, the probability of channel blockage is calculated and the energy consumption per bit is estimated. Treating carotenoid molecules as transmitters, chlorophylls as receivers, and the energy transfer between them as a way to encode information, a throughput up to 1 Gbit/s is achievable with a bit error rate below $10^{-3}$, average transmission delays about 20 ps, and energy consumption c.a. $2.0 \times 10^{-18}$ J/bit. These results indicate a high potential of photosynthetic systems for nanocommunications and other related applications, due to suitable energetic characteristics in terms of energy harvesting abilities and low consumption for data transmissions.

## I. INTRODUCTION

Nanocommunications, intensively discussed in biological and engineering research societies, are commonly understood as a means for providing connectivity between various nanomachines. Currently, the main approach in nanocommunications is related to processes and phenomena where information is carried by molecules, vesicles or even bacteria in natural environments of living organisms, which is therefore called molecular communications [1]. With the advantages of bio-compatibility and energy efficiency, these molecular mechanisms perform very well in numerous applications, especially those related to human body, for example, nanomedicine [2]. The medical applications pertain to both diagnosis and treatment of diseases [3]. Diagnosis can concern disease detection, e.g. via detection of specific bacteria, cytokines or bio-markers concentration in blood [4, 5], as well as imaging techniques, e.g. with fluorescent molecules where Förster resonance energy transfer (FRET) can be used both for communication and for molecular and cellular imaging [6]. On the other hand, molecular and nano-level approaches offer new opportunities for treatment of diseases as well, through targeted drug delivery [2], tissues engineering [7, 8], nanosurgery [9] and immune system triggering [10]. Applications from other fields related to data storage in molecules, brain-computer interactions and connecting nano-systems with larger-scale communication networks [11, 12] are arising as well.

In this paper, photosynthetically active pigments are introduced for the purpose of nanocommunications. They are very well known in biological sciences, as being responsible for harvesting sunlight energy, which is necessary for conducting photosynthesis, a process of primary significance to all living organisms. As commonly occurring in plants, algae and cyanobacteria, they are bio-compatible with many other biological molecules and structures. However, the potential of photosynthetic pigments in communications has not been recognized. Numerous unique abilities make them extremely promising in the perspective of nanocommunications. Hence, in the further parts of the paper, particularly Section II and V, the most significant features of photosynthetic pigments will be highlighted:

a) as naturally created for this purpose, they are very effective in harvesting sunlight energy, which can also provide the source of energy for communications;
b) they are very energy-efficient when sending signals between each other, what corresponds to extremely low energy-per-bit values;
c) they operate with very small delays, releasing energy and passing signals in picoseconds;
d) they are able to convert signals from one type to another: accepting photons, communicating via Förster resonance energy transfer and releasing electrons. Photosynthetic pigments communicate between themselves using FRET. In such communications, the transmitter and the receiver are fluorescent molecules, being able to absorb and emit energy in the visible part of electromagnetic spectrum. The information to be transmitted is encoded in the energy transferred between the molecules, without emission of any photons or chemical signals.

Focusing on photosynthetic pigments, the scientific contribution of this paper is as follows. We introduce photosynthetic pigments, highlighting their characteristics suitable for nanocommunications. To support these theoretical considerations, two experiments focusing on the energy transfer between photosynthetic pigments and its efficiency are conducted and reported in this paper. These experiments do not realize a full communication system with data coding and modulation, which is still an open research issue. However, the obtained results serve us a basis for calculations of the communication bit error rate and throughput. The experiments are performed on a well-known structure, photosystem II (PSII) complex, containing photosynthetic molecules such as chlorophylls and carotenoids communicating via FRET. While FRET has been already studied for nanocommunications in some artificial, non-photosynthetic systems (AlexaFluor, DyLight or GFP proteins [13, 14]), here we provide evidence that such a transfer between carotenoids and chlorophylls is about 2 orders of magnitude faster than already reported [13, 14] and about 30 times more energy-efficient than reported in [15].

The presented paper is organized as follows. In Section II, the photosynthetic structures, chlorophylls and carotenoids as two main families of photosynthetic pigments are introduced. In Section III, the basics of FRET which is the phenomenon utilized by carotenoids and chlorophylls for information transfer are delivered. We also comment on coherent resonance energy transfer (CRET) and Dexter which are two other possible phenomena for energy transfer between molecules. In Section IV, a brief description of both laboratory experiments performed on the PSII system is provided. In Section V, the communication performance of the carotenoid→chlorophyll transfer, focusing on bit error rate, throughput and energy consumption is thoroughly analyzed. Future research challenges and open issues are summarized in Section VI. Finally, in Section VII, we conclude the paper.

## II. PHOTOSYNTHETIC STRUCTURES

Among numerous bio-molecules and bio-structures that can be considered in molecular communications, the photosynthetic ones have very unique advantage: they are able to harvest sunlight energy further used for driving photosynthesis. As designed by nature in a long evolution, they are very efficient in this task. The energy transfer within photosynthetic structures proceeds through a cascade of photosynthetic pigments, the resonators, among which the excitation energy could fluctuate before being trapped or stabilized. After the absorption of light, such pigments (photosynthetic antennas) cooperate in the transmission of photon flux to the reaction center, where finally, the excitation is trapped. When applied on surfaces exposed to sunlight, photosynthetic structures could be a source of energy for other nanomachines or larger systems. They could also be considered as parts of nanostructures for numerous kinds of intelligent fabrics and materials. Furthermore, they are also extremely fast in forwarding FRET signals, as their exponential lifetimes are on the order not of nanoseconds (like in case of synthetic dyes), but pico- or even femtoseconds [16].

The two major groups of photosynthetic molecules that usually occur together, are chlorophylls and carotenoids. Chlorophylls are the most abundant natural pigments responsible for green coloration of plants. They bind to special proteins, enabling formation and contributing to the stability of so produced photosynthetic structures, namely light-harvesting complexes type I and II (LHCI and LHCII), and type I and II photosystem reaction centers (so called PSI and PSII, respectively). Chlorophylls and their derivatives are capable of visible light absorption roughly in the range of ~330 to 800 nm, which is the most intensive in the violet (<440 nm) and red (>650 nm) part of the spectrum. Chlorophylls' functions are predominantly linked to photosynthesis. They absorb light and transfer the excitation energy with extremely high (~100%) quantum efficiencies from LHC to reaction centers, where the primary charge separation takes place, followed by electron transportation via photosynthetic electron transport chain [17].

In contrary to chlorophylls, carotenoids constitute a group of pigments much more diverse in terms of structure [18]. They fulfill a number of functions in a wide range of biological systems [19]. In photosynthesis, their main roles are light energy harvesting and photo-protection. Hence, they act as accessory pigments gathering light energy from the visible spectrum regions (violet-blue-green part) that are not fully accessible for chlorophylls. In consequence, they pass, via

FRET, most of the captured energy to chlorophylls. The remaining part is dissipated as heat, as carotenoids emit nearly no light themselves (their fluorescence is insignificant with quantum yield below 0.01%, i.e., less than 1 out of 10000 absorbed photons are emitted) [20]. Carotenoids are also very efficient physical and chemical quenchers of singlet oxygen as well as scavengers of other reactive oxygen species [17, 21], protecting in this way all photosynthetic pigment-protein structures from irreversible modification and degradation. Apart from their role in photosynthesis, both types of pigments may interact with other biological structures and their components like, for example lipids and polypeptides (proteins) forming membranes, or low-molecular weight compounds like cytochromes or vitamins.

In the PSII complex, chlorophylls and carotenoids cooperate efficiently in harvesting sunlight energy. It might be judged from their molar absorption coefficients $\varepsilon$, which are one of the highest amid natural compounds [20]. For synthetic dyes, such as for example Alexa Fluor 488 (roughly comparable to β-carotene with its spectral range), $\varepsilon$ is also significantly lower [22].

Another important feature of photosynthetic molecules is their swiftness in forwarding signals (releasing energy). When β-carotene, a component of PSII absorbs photon, it gets excited and its energy level goes from the ground-state $S_0$ directly to $S_2$ state (Fig. 1). Recent studies revealed, that the surprising lack of $S_0$ to $S_1$ transition is not due to β-carotene symmetry properties as was thought for a long time, but rather it must have another physical origin, such as deformation of the molecule in the low-lying $S_1$ state [23]. The excited $S_2$ state can very quickly decay to the state $S_1$ (internal conversion) or to $S_0$ (via FRET or heat dissipation); this decay lasts 100-300 fs, on average [24]. From the state $S_1$, the molecule undergoes the transition to $S_0$ in about 9 ps [25], also via FRET or heat dissipation. Hence, the whole process of releasing energy lasts no longer than 9.3 ps on average, but it may vary, as the excited state lifetimes are given by exponential distributions. Thereby, both the $S_1$ and $S_2$ states function as energy donors, however the transfer from the $S_1$ state was shown to be less favored [16].

On the other hand, a chlorophyll-a molecule, after absorbing energy from sunlight or via FRET from a β-carotene, contributes to the initial reactions of photosynthesis. An excited chlorophyll-a starts this process by releasing an electron and returning to the ground-state in about 10 ps [26]. This electron usually activates next steps of photosynthesis, but it can be caught by other devices as well. Thus, molecules such as chlorophyll-a can be thought of as FRET-to-electron converters. Comparing with synthetic fluorescent dyes like AlexaFluor, DyLight or GFP, where excited state lifetimes are measured in nanoseconds, usually 0.3-4 ns [14, 27, 28], photosynthetic molecules are about 2 orders of magnitude faster in emitting their energy. It has strong implications on the possible data throughput and will be touched upon in Section V.

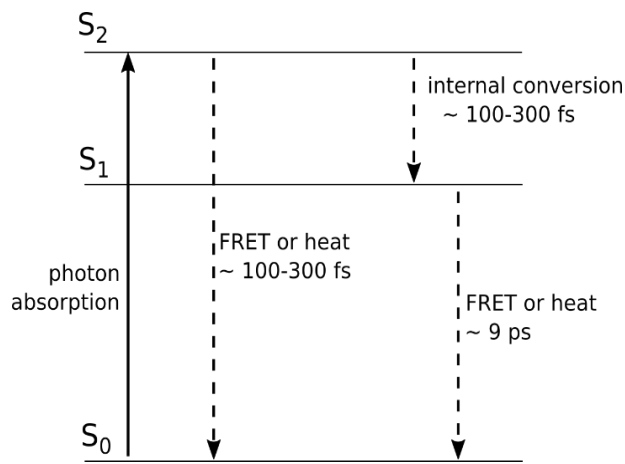

**FIGURE 1.** Simplified energy level diagram for β-carotene.

### III. RESONANCE ENERGY TRANSFER

Förster resonance energy transfer is a fundamental mechanism of the energy transfer between photosynthetic molecules and thus it will be reviewed in more details in this section. While both chlorophyll and carotenoid molecules harvest energy directly from sunlight, in different regions of the visible spectrum, the chlorophylls receive it also via FRET from carotenoids. FRET is a phenomenon where an excited molecule, called a donor, non-radiatively passes its energy to another molecule, called acceptor [27]. Both molecules must be located close to each other, usually up to 10 nm. Also, they must be spectrally matched, i.e., the donor emission spectrum should, at least partially, overlap the acceptor absorption spectrum (there are however, some exceptions to this rule when the donor excitation energy is increased with so called phonon-assisted energy transfer [29]). The donor molecule can be excited in many ways, e.g., accepting a photon or using energy of a chemical reaction. After excitation, the donor keeps its energy a certain time; this delay time is exponentially distributed with the average value depending on the molecule type (molecule excitation times are commonly modeled with exponential distributions [27, 30]). Then, the energy transfer between the donor and the acceptor is realized without any radiation and the donor is again in its ground, i.e., not excited, state. The FRET efficiency ($E$) is highly dependent on the separation between the donor and acceptor molecules, it was shown to be equal to [27]:

$$E = \frac{R_0^6}{r^6 + R_0^6} \quad (1)$$

The parameter $r$ is the separation between the donor and acceptor and $R_0$ is the so called Förster distance, which depends on the spectral match between these two molecules. The value $R_0$ can be determined experimentally or calculated having known the donor quantum yield, donor emission spectrum and acceptor emission spectrum [27]:

$$R_0^6 = 0.211 \cdot \kappa^2 n_r^{-4} Q_D \int F_D(\lambda) \varepsilon_A(\lambda) \lambda^4 d\lambda \quad (2)$$

where $F_D(\lambda)$ is the donor emission spectrum, $\varepsilon_A(\lambda)$ is the acceptor molar absorption coefficient in the function of wavelength, $Q_D$ is the quantum yield of the donor molecule in the absence of the acceptor molecule, $n_r$ is the refractive index and $\kappa^2$ is a parameter describing a relative orientation of donor and acceptor dipoles. Depending on the molecule type, $R_0$ typically varies from 2 to 9 nm [27, 28].

For the communication purposes, the energy transfer between a donor and an acceptor can be treated as a way to send information bits over a channel consisting of a physical space between these two molecules. The donor plays the role of a nanotransmitter and the acceptor is the nanoreceiver [31]. The transmission of information is realized when the excitation energy is passed via FRET from the donor to the acceptor, so after the transmission the donor is in the ground state and the acceptor is in the excited state. In such a scenario, it has been already proposed to use ON-OFF modulation [32]. Keeping a common assumption of a kind of synchronization between the transmitter and receiver sides, a bit '1' is sent exciting the donor and expecting the FRET occurs, while a bit '0' is sent keeping the donor in the ground state (low energy level) so that no energy reaches the acceptor (no FRET). While the issue of donor excitation technique is out of scope of this article, it should be noted that both internal option, e.g., via bioluminescence or another FRET, as well as external one with optical filters mentioned in Section II, are feasible.

Equation (1) shows that, especially for the distances larger than $R_0$, FRET is very inefficient. The energy, instead of being passed to the acceptor, is emitted in the form of a photon (in case of fluorescent molecules) or dissipated as heat. From the communication viewpoint, such energy is lost and a transmission error occurs. Assuming no hardware noise and a correct transmitter-receiver synchronization, we adapt so called Z-channel model [29], where errors do not happen when sending a '0', as there is no respective energy transfer. Errors may happen only when transmitting bits '1', with the probability $1-E$. Presuming, on average, 50% of bits are 'ones', the corresponding bit error rate (BER) is given as:

$$BER = 0.5(1-E) \quad (3)$$

as $E$ gives the probability of a successful energy transmission. Even when the donor and acceptors molecules are separated by about Förster distance, the related BER is clearly not acceptable for the communication (usually the required BER is $10^{-3}$ or better). This is a reason why the Multiple-Input Multiple-Output FRET (MIMO-FRET) technique was proposed, where multiple donors and acceptors are used at the same time, analogous to MIMO wireless systems, in order to enhance the reliability of the signal transfer. It was already shown that having $n$ donors and $m$ acceptors, the related BER is much smaller than in the case with single molecules and can be expressed as [14]:

$$BER_{n,m} = 0.5 \cdot \left( \frac{r^6}{r^6 + m \cdot R_0^6} \right)^n \quad (4)$$

Having 5-6 molecules both at transmitter and receiver sides enables BER below $10^{-3}$ for the communication distances equal to the Förster distance [13, 14].

FRET is a phenomenon being in interest of communication researchers since [33], where it was mentioned in the context of interfaces between molecular communication and optical signaling. Later, a FRET-based physical channel was proposed in [32] with the multi-hop communication analyzed in [34]. Some measurement works appeared after, where the communications were studied between popular fluorescent dyes, like Fluorescein and Rhodamine [35], Alexa Fluor [13] and DyLight molecules [14]. These works reported large possible throughput of FRET-based nano-networks, but also their limited communication range and high bit error rate, thus MIMO-FRET was suggested. Routing techniques for FRET-based networks were also proposed in [36] and [31], mainly exploiting some options of chemical manipulation of the shape, movement and emission properties of fluorescent molecules. A broad overview of FRET-based communications can be found in [37] with the focus on networking and architecture aspects, as well as on applications. Also, interesting interfacing techniques between fluorescent molecules with bioluminescent stimulation and some larger medical systems are discussed in [38] and [15], connecting FRET-based networks with neurons as well [39]. A broad discussion on interfaces between FRET-type communications and other networks can be found in [12].

Other research concepts presented recently, may also help to increase the communication range via FRET on specific materials. The first one is vibrating FRET (VFRET), which is the energy transfer between molecules located on a vibrating multi-layer graphene membrane [40]. The effective range of this mechanism can reach 20 nm. It is also reported that FRET can occur on distances about 160 nm [41], which however requires using an optical topological transition in a metamaterial. In both these cases, the donor molecules are quantum dots.

For the completeness of the above description, other mechanisms of energy transfer between molecules, like coherent resonance and Dexter, should be mentioned here; however, they do not occur in the experiments reported in this paper. CRET occurs if the coupling between involved molecules is much stronger than for FRET. Theoretical investigation of this phenomenon shows that CRET is possible in ambient temperatures [42]. Its evidence, however, is practically confirmed only in cryogenic conditions (77 K) [43]. Finally, Dexter energy transfer is another non-radiative process where two molecules exchange excited electrons. Unlike FRET, its efficiency decays exponentially with the distance between molecules and usually occurs for very short distances up to 1 nm [44], which is not the case of the scenarios considered here.

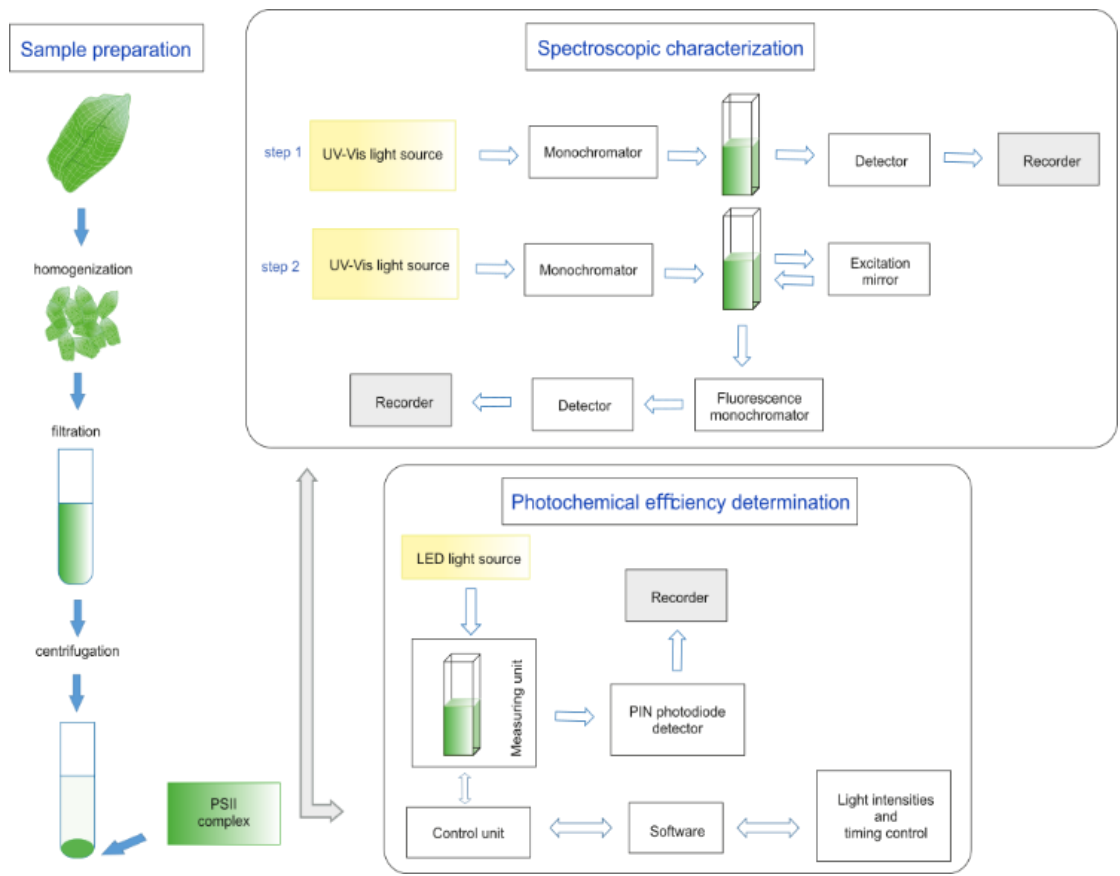

**FIGURE 3.** Left side: major steps of sample preparation. Right side: schematic presentation of setups applied for experiments, detection of absorbance and steady-state fluorescence spectra (upper part) and measurement of fluorescence kinetic curves (bottom part).

## IV. EXPERIMENTAL PART: DETERMINATION OF ENERGY TRANSFER AND PHOTOCHEMICAL EFFICIENCIES

The two experiments focused on PSII complex, were meant to supplement theoretical considerations on utility of photosynthetic pigments for communication purposes. A PSII complex, is a $21\times11\times10$ nm$^3$ molecular structure spanning across the membrane with water oxidizing complex facing the lumen (interior of thylakoids). PSII is a homodimer with a two-fold symmetry axis running through the membrane plane [45]. Each monomer is composed of proteins and cofactors including pigments (carotenoids and chlorophylls) located in a close proximity to each other. As reported [46], FRET can occur between them, with β-carotenes being donors and chlorophyll-a molecules serving as acceptors (Fig. 2).

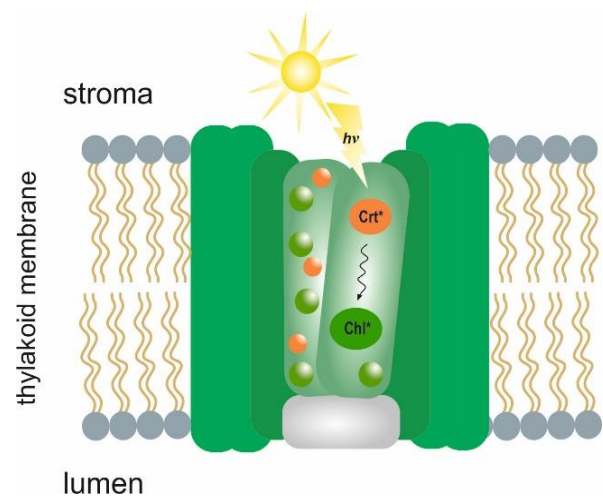

**FIGURE 2.** Simplified presentation of photosystem II complex (PSII). Only the core part of PSII comprising conserved transmembrane proteins (green blocks) and some of the pigments as orange (carotenoids) and green (chlorophylls) circles are shown. Used abbreviations: Crt – carotenoid molecule; Chl – chlorophyll molecule; lumen/stroma – the space inside/outside the thylakoid membrane (photosynthetic membrane).

The first experiment was focused on the determination of the efficiency of the carotenoid→chlorophyll energy transfer, in order to show that signals can be transmitted from one type of molecules to the others. However, one has to keep in mind that calculated values refer to an average of many carotenoid-chlorophyll pairs that are present in a model PSII complex. The aim was also to check the photosynthetic activity of PSII under the action of light and the usability of sample for further experiments. The goal of the second experiment was to measure the energy releasing from chlorophylls, which enabled analysis of the whole energetic process related to the FRET transfer.

### A. PLANT MATERIAL

Thylakoids, i.e., photosynthetic membranes containing PSII complexes, were isolated from fresh leaves of spinach (*Spinacia oleracea* L.) according to the method described in [47] with minor modifications. The major steps of the preparation procedure are summarized in Fig. 3 (left side) and described in more detail in Appendix.

### B. SPECTROSCOPIC CHARACTERIZATION OF PSII

In order to determine the FRET efficiency between carotenoids, (β-carotene), and chlorophylls (chlorophyll-a), a series of absorption and fluorescence spectra of the PSII system were recorded. Steady-state absorption spectra were measured with a Cary50 Bio UV-VIS spectrophotometer (Varian) in a standard quartz cell in the range of 300-850 nm. Steady-state fluorescence (excitation and emission) spectra were measured on a CaryEclipse spectrofluorometer (Varian). Emission spectra were recorded within a range of 650-800 nm upon excitation at 475 nm, 485 nm, or 500 nm (carotenoid band), and 438 nm (chlorophyll band), respectively. Fluorescence excitation spectra were collected in the range of 350-600 nm, while emission from chlorophyll was measured at 684 nm. The layout of the key components of both instruments are schematically presented in Fig. 3.

The FRET can be observed from β-carotene to chlorophyll-a, as the β-carotene emission spectrum and the chlorophyll-a absorption spectrum overlap for the wavelengths about 500 nm (β-carotene absorption and emission spectra are strongly overlapped, as the respective Stokes shift is very small). The fluorescence of β-carotene molecules is known to be very low [16, 20], so the FRET efficiency cannot be measured directly by estimating the decrease of the donors (β-carotene) lifetime in the presence of the acceptors (chlorophyll-a), like in [13, 14]. Instead, it was decided to use another approach as presented in [48], and further described in details in [49, 50]. It is based on comparing two respective spectra. The first one is the *1-T(λ)* spectrum, where *T(λ)* is the transmittance calculated on the basis of the absorption spectrum as: $Ab(\lambda)=-\log_{10}T(\lambda)$. The second one is the *Fl(λ)* spectrum (fluorescence excitation). Thus, the fluorescence excitation shows how much energy was absorbed and passed through the PSII system. Both spectra, *1-T(λ)* and *Fl(λ)*, can be normalized to each other, as it is known that their values at 600 nm must be equal (at 600 nm predominantly chlorophyll-a absorbs energy).

### C. SPECTRA ANALYSIS AND FRET CALCULATION

The PSII absorption, excitation and emission spectra were recorded 20 times. The first four measurements were rejected since they might not be fully reliable due to the sample stabilization. The final spectra were proceeded as the averages of the remaining 16 records, although the differences between them were negligible. In Fig. 4 the absorption spectrum of PSII and the emission of chlorophyll-a upon excitation of PSII at carotenoid band are presented.

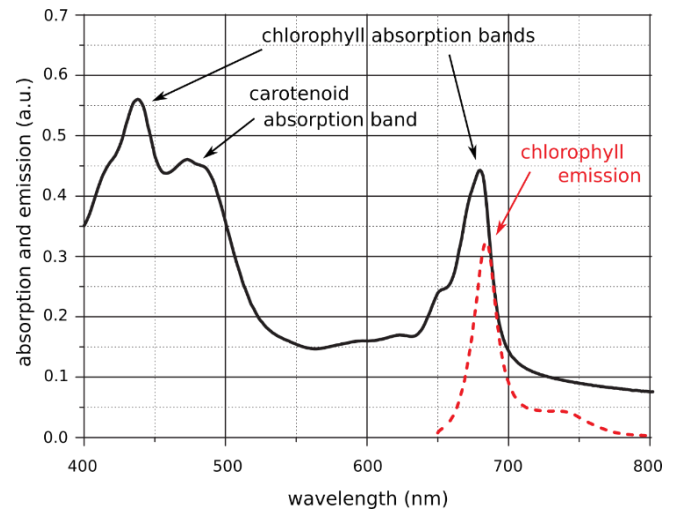

**FIGURE 4.** The PSII absorption spectrum (black solid line) and emission of chlorophyll upon excitation of PSII at 500 nm (red dash line).

One can clearly notice the absorption maxima related to the presence of carotenoids and chlorophylls. The chlorophyll-a emission spectrum was obtained upon PSII excitation at 500 nm. Its presence corroborates the activity of the components of PSII complex.

In Fig. 5, the *1-T(λ)* (calculated on the basis of the absorption) and the *Fl(λ)* (fluorescence excitation) spectra are normalized at 600 nm. In the experiments, the FRET efficiency was calculated as the *Fl(λ)/(1-T(λ))* ratio for wavelengths from 480 to 500 nm, which is the sole carotenoid absorption range. Under our experimental conditions, the estimated ratio varied from 84.4% for $\lambda = 480$ nm to 64.0% for $\lambda = 500$ nm. The final FRET efficiency was obtained by averaging the ratio over the whole 480–500 nm spectral range and equaled 75.5±1% (based on the instruments uncertainty). The obtained result can be compared with theoretical values. In the PSII system, the physical distance between β-carotene and chlorophyll-a molecules is about 1.8 nm [51], so if the FRET efficiency is 75.5%, the Förster distance should be about 2.17 nm (see (1)). And indeed, $R_0$ for β-carotene – chlorophyll-a pair is assessed as 1.8–2.5 nm [52].

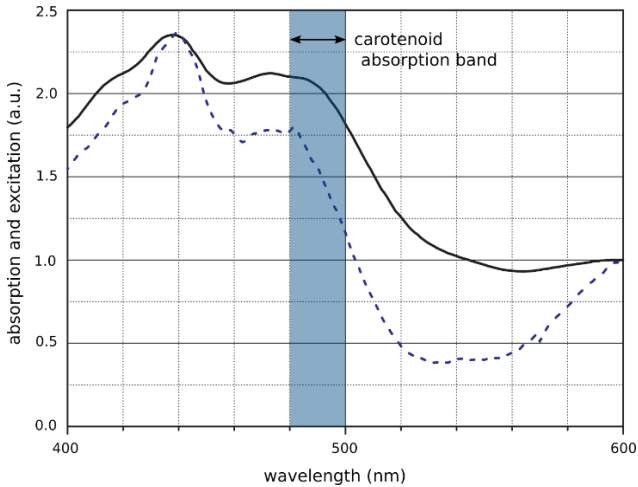

**FIGURE 5.** 1-T(λ) (black solid line) and Fl(λ) (blue dash line) spectra, of the carotenoid range, normalized at 600 nm.

### D. PHOTOCHEMICAL EFFICIENCY

The goal of the second experiment was to confirm the chlorophyll-a photosynthetic properties and measure the efficiency of the primary photochemistry of photosynthesis. For this purpose, chlorophyll-a fluorescence kinetic curves were recorded. The origin of a fluorescence curve represents a minimum value, i.e. the initial chlorophyll fluorescence ($F_0$). Under saturating excitation light the value of fluorescence reaches its maximum (maximal fluorescence yield, $F_M$). The difference between $F_M$ and $F_0$ (variable fluorescence, $F_V$) expressed as the ratio of $F_V/F_M$ defines the photochemical quantum efficiency $\Phi$.

The experiment was performed at room temperature using double modulation fluorometer FL 3300 (Photon Systems Instruments). Fast fluorescence kinetics were measured on 1 s time scale: (a) up to 2 ms with the 10 µs step and (b) from 2 ms with the 1 ms step. Following a 20 min dark adaptation, samples were illuminated with continuous red light (625 nm) provided by LEDs and chlorophyll fluorescence signals were recorded according to [53]. The maximal PSII photochemical efficiency in the dark-adapted state ($F_V/F_M$) was calculated following the equation:

$$\Phi = \frac{F_V}{F_M} = \frac{F_M - F_0}{F_M} \quad (5)$$

An example of the measured curve is presented in Fig. 6. The fast phase (up to 1 s) of chlorophyll-a fluorescence kinetic curve was measured 5 times. The obtained $F_0$ and $F_M$ chlorophyll-a fluorescence levels were used to calculate the photochemical efficiency according to (5) and equaled 83.6±1.5%. This stays in a good agreement with other similar measurements reported in literature [46, 54, 55].

### V. COMMUNICATION PERFORMANCE

In this section, the literature data and the results of the reported experiments are used to evaluate the usefulness of photosynthetic pigments, carotenoids and chlorophylls, for nanocommunications. In the respective subsections, bit error rate, throughput and energy consumption are analyzed.

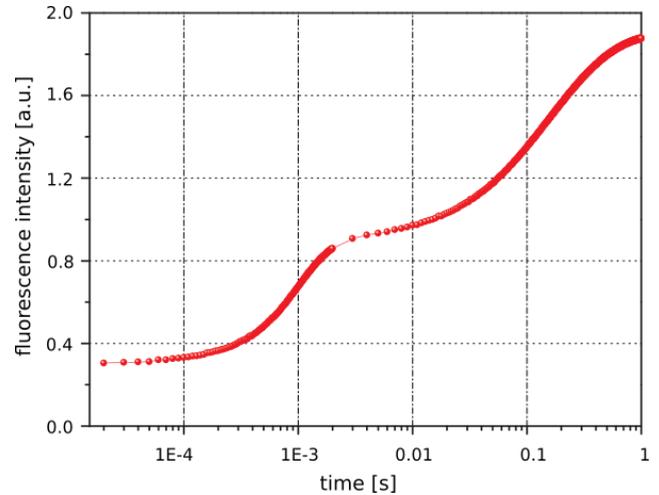

**FIGURE 6.** Chlorophyll-a fluorescence kinetic curve measured for the PSII complex.

### A. BIT ERROR RATE

The FRET efficiency of 75.5% measured and reported in Section IV is quite high comparing with other experimental results [13, 14], but the respective BER (12%, see (3)) is still not acceptable for communications. A viable solution can be repetition of FRET transmission for each bit $p$ times, then assuming a correct reception of a bit '1' if at least one of the FRET transmissions is correctly received. Thus, an error occurs only if a bit '1' is transmitted and all $p$ transmissions fail. This is caused by the fact that the communication channel is not symmetric, i.e., bit errors can happen only when transmitting '1', and it further leads to the modification of (3) into:

$$BER = 0.5(1 - E)^p \quad (6)$$

In Fig. 7, the BER curves are given for FRET transmissions with bit repetitions 1 to 10 times.

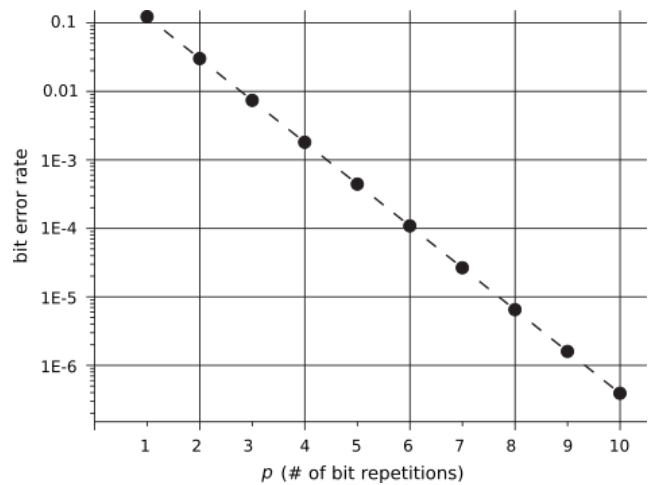

**FIGURE 7.** Bit error rate for FRET transmission in the PSII system as a function of number of repeating bits.

Bit error rates below $10^{-3}$ can be obtained for $p = 5$, while for $p = 10$, a very good value of BER below $10^{-6}$ is achievable.

## B. THROUGHPUT

As discussed in [14], the throughput of a FRET communication channel depends on the exponentially distributed lifetimes of both donor (carotene) and acceptor (chlorophyll), as these values tell us how quickly energy is released by the molecules. Assuming, for a while, no bit repetitions, if the communication is carried with the throughput of $C$ bits/s, then the transmission of a single bit should be realized in the time of $T=1/C$. In that time, the donor must return to its ground state (it must release the excitation energy via FRET or photon) in order to be able to accept another quantum of energy. Also, the acceptor should release its energy, e.g., to a decoding device (which is not specified in this paper) or another communication node. Then, both the donor and acceptor molecules are ready for another FRET transmission. Because of the exponential lifetime distributions [27, 30], the probability of donor or acceptor being blocked by a previous excitation is never equal to 0. One may, however, calculate the probability that, at first, the donor and then, the acceptor will pass the signal and both these events will happen consecutively in the time $T$ [s]. First, let us assume the donor signal transfer happened at time $t$ [s]. Then, the acceptor must release the signal in the time $T-t$, which is given by the cumulative distributed function (cdf) of its lifetime exponential distribution as:

$$P_A(T-t) = 1 - e^{\frac{t-T}{\tau_A}} \quad (7)$$

where $\tau_A$ is the average acceptor lifetime. Now, we can define the probability of the both (donor and acceptor) signal transfers happening consecutively, using an integral over the period $(0,T)$ of the donor lifetime probability density distribution, with the acceptor cdf as a parameter:

$$P_{DA} = \int_0^T \frac{1}{\tau_D} e^{\frac{-t}{\tau_D}} \left(1 - e^{\frac{t-T}{\tau_A}}\right) dt \quad (8)$$

where $\tau_D$ is the average donor lifetime. The integral from (8) may be solved as:

$$P_{DA}(T) = 1 - \frac{\tau_A e^{\frac{-T}{\tau_A}} - \tau_D e^{\frac{-T}{\tau_D}}}{\tau_A - \tau_D} \quad , \text{ for } \tau_A \neq \tau_D \quad (9a)$$

$$P_{DA}(T) = 1 - e^{\frac{-T}{\tau_A}} - \frac{T}{\tau_A} \cdot e^{\frac{-T}{\tau_A}} \quad , \text{ for } \tau_A = \tau_D \quad (9b)$$

The probability $P_{DA}(T)$ in (9b) is given by the Erlang distribution which is the special case of the gamma distribution in (9a) [56]. At the same time, (9b) can be derived from (9a) using the L'Hôpital's rule.

Now, we can finally define the probability that the FRET channel is blocked because the donor or the acceptor being still in the excited state; it is $P_{Blocked}(T) = 1 - P_{DA}(T)$ which can be further expressed as:

$$P_{Blocked}(T) = \frac{\tau_A e^{\frac{-T}{\tau_A}} - \tau_D e^{\frac{-T}{\tau_D}}}{\tau_A - \tau_D} \quad , \text{ for } \tau_A \neq \tau_D \quad (10a)$$

$$P_{Blocked}(T) = e^{\frac{-T}{\tau_A}} + \frac{T}{\tau_A} \cdot e^{\frac{-T}{\tau_A}} \quad , \text{ for } \tau_A = \tau_D \quad (10b)$$

For the PSII system investigated in this paper, the donor and acceptor lifetimes are known from the literature analysis, and they are about 9.3 ps for the donor, β-carotene, and 10 ps for the acceptor, chlorophyll-a (see Section II). In Fig. 8, $P_{Blocked}(T)$ is plotted for these values. We can interpret this probability as a factor that additionally increases the BER for a single bit transmission. For $T>100$ ps, $P_{Blocked} < 10^{-3}$, which is insignificant comparing with BER of 0.12 (see Section V.A). The corresponding throughput $C$ is 10 Gbit/s. With the bit repetition applied (see the previous section), the throughput is still above 1 Gbit/s, which is very high for such a nanosystem. We can conclude that the FRET channel blockage is not a limiting factor and the real system throughput depends rather on the external source of signal and hardware limitations, like thermal noise, timing distortions and synchronization issues between the transmitter and the receiver sides. These problems are well illustrated in [35].

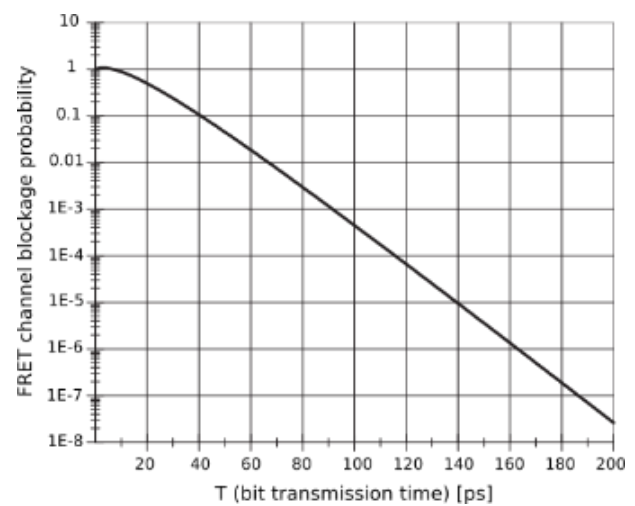

**FIGURE 8.** The FRET channel blockage probability being the result of the donor or acceptor still in excited state.

## C. ENERGY CONSUMPTION

Photosynthetic molecules like carotenoids and chlorophylls, have great advantages of not only bio-compatibility, but also high energy efficiency and energy harvesting capability. Discussing their behavior from the energetic perspective, several points should be highlighted.

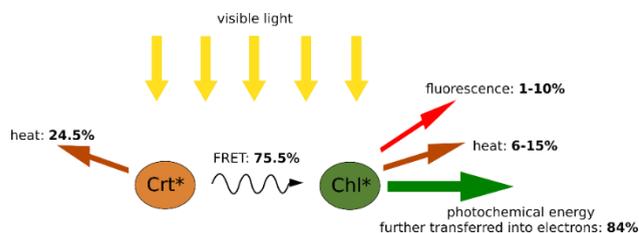

**FIGURE 9.** Schematic representation of the energy flow within the PSII complex based both on the data obtained in the present study (FRET and photochemistry) and literature. Labels: Chl*, Crt* – excited state of chlorophyll and carotenoid, respectively.

First, both types of pigments can harvest energy directly from solar radiation, on different sub-bands of visible light spectrum: 400-550 nm for carotenoids (violet-blue-green region), ~400-450 and 640-700 nm for chlorophylls (violet-blue and red regions, respectively). Thus, when they are exposed on daylight, communications between them can be realized independently, without any extra energy supply. Additionally, the intensity of this energy harvesting and communications can be tuned via optical filters or nanoparticles, adjusting amount of incident light. Carotenoids emit FRET signals that are further converted into electrons by chlorophylls. Additionally, chlorophylls can harvest more energy if needed, under control of another optical filter.

Second, nanocommunications via FRET are highly energy efficient. The energy spent for a single FRET transmission is equivalent to the energy of a photon, as these are two different ways how an excited molecule may return to its ground state. Thus, this energy can be assessed with the Planck-Einstein relation (Energy = $h \times$ frequency, where $h$ is a Planck constant, $h = 6.626 \times 10^{-34}$ J·s). For the wavelength of 500 nm, where β-carotene emission and chlorophyll-a absorption spectra strongly overlap, the energy is about $4.0 \times 10^{-19}$ J. With ON-OFF modulation (see Section III), only bits '1' require FRET, so assuming the same number of 'ones' and 'zeros', the energy should be further divided by two. On the other hand, for reasonable BER we proposed bit repetitions even 10 times, so finally we may assess the energy of a single bit as $2.0 \times 10^{-18}$ J. This value is few orders of magnitude smaller than in the case of other molecular communications, where energy per bit is usually between $10^{-16}$ and $10^{-13}$ J [57, 58], and also 30 times smaller than for another FRET-based communication techniques reported so far [15]. It should be noted, however, that FRET is clearly a short-range communication mechanism comparing with other molecular options.

Finally, according to literature [46, 59], partitioning of absorbed excitation energy in assemblies containing carotenoids and chlorophylls, can be utilized through one of the three competing pathways: (1) photochemistry (primary charge separation and photosynthetic electron transfer), (2) thermal dissipation or (3) fluorescence emission. The results of the second experiment reported in this paper (Section IV.B) delivered more information on how the energy is distributed within the analyzed system. The most important data, which is the efficiency of a primary PSII photochemistry, was determined as about 84%. This result indicates that chlorophylls, major constituents of PSII, are very effective in processing FRET energy, which is later converted into electrons [46]. The remaining 16% of energy is spent in different ways, out of which chlorophylls' fluorescence may reach up to 10%, while the remaining part is emitted as heat [60, 61]. The schematic energy flow within the PSII system is shown in Fig. 9.

## VI. OPEN ISSUES

Both, experiments and theoretical considerations, provided a broad view on how and at what cost signals can be transmitted in a complex system built on photosynthetic molecules. The energy efficiency and harvesting features, as well as the ability of fast signal conversion look very promising for future generations of nanodevices based on bio-molecules. Still, some issues remain open and should be elucidated when thinking about future research and applying such structures in molecular communication systems. These are, especially, data transmission between single photosynthetic molecules, photosynthetic pigments stability, integration of pigments in nanomachines, control on energy harvesting and electron detection for data transmission, which will be discussed in the following paragraphs with suggestions on further directions of studies.

a) Data transmission between single photosynthetic molecules - the experiments reported in the present paper were focused on energy transfer in a sample containing a large number of PSII complexes, having in each monomer 35 chlorophylls and 11 β-carotenes [62]. Further studies should involve a real data transmission, including coding and modulation between single molecules: one donor and one acceptor for the FRET-based communication. Such scenario could be realized with a single molecule FRET fluorescence microscopy, a technique well-known in biology [63, 64], but still not recognized for communication studies.

b) Photosynthetic pigments stability - these pigments are very robust only in their native environment. If not bound to other bio-molecules, they are not as stable as their synthetic counterparts and easily undergo degradation in the presence of light and oxygen. Here, one possible solution could be their structural modification. The stability of chlorophylls can be increased by modification of the porphyrin ring, central metal ion substitution, addition or removal of peripheral substituents or aggregation [65, 66]. The modification of carotenoids may include cyclisation of one or both ends, attachment of oxygen groups and heteroatoms, generation of aggregates [67]. Another possibility is a controlled assembly of photosynthetic structures by incorporation of desired pigments, polypeptides and other molecules (nanoparticles, nanowires) to the reconstituted complexes, or even use of genetically modified proteins with specifically altered properties. Such procedure may result in formation of artificial photosynthetic structures of improved optical functionality, that might be further tuned in order to fulfill the severe criteria of nanocommunications [50, 68].

c) Integration of pigments into future nanomachines - photosynthetic pigments could be considered as parts of nanostructures for intelligent fabrics and materials, but it is still not fully clear how to use intact pigments and their protein complexes as building blocks of nanodevice systems. To achieve this, various approaches are tested, as retaining their unique biological activity is one of the most challenging issues. Hence, there are examples of functional hybrid complexes comprising liposomes and photosynthetic structures, their combinations with carbon-based materials or transition metal oxides [69].

d) Control on energy harvesting - when applied on surfaces exposed to sunlight, photosynthetic pigments can be a source of energy for nanomachines. The process of energy harvesting should be controlled and it can be realized by both internal (e.g., cytochrome $c_{553}$) as well as numerous external filters, activating strictly desired chlorophyll's pools [70]. The external option might be an optical thin-film resonant cavity filter based on a Fabry-Perot or Mach-Zehnder interferometer [71]. Other possibilities are nanoparticles or nanowires filtering the energy gathered by light harvesting complexes [72, 73]. All these filters can be used for the modulation of the transferred signals or control over the functions of nanomachines or other hybrid nanostructures [74]. Furthermore, due to the exceptional spectroscopic properties of chlorophylls, such as their long-lasting triplet states [75], as well as the possibility of various structural modifications [65] they turned out to be potent "therapeutic nanoagents" in photomedicine [76]. Photosynthetic pigments bound to proteins and attached to different carrier matrices might also serve as promising bio-nanocomposite materials used in a broad range of optical devices, e.g., optical switches, micro-imaging systems, (bio)sensors etc. as well as be used in a variety of other optoelectronic applications [69].

e) Electron detection for data transmission – the fact that chlorophylls convert FRET signals into electrons is both a great opportunity and a research challenge. First, it creates a chance to store a FRET signal and further re-send it to other networks. On the other hand, a proper interface technology intercepting electrons is required. It can be realized using electron ptychography and direct electron detectors [77, 78] developed in recent years and achieving ångström-level resolution. These techniques could connect photosynthetic systems with electronic equipment working in micro or larger scale. The second possible solution could be connecting photosynthetic pigments with other molecular communication systems via redox (reduction-oxidation) reactions. These reactions are ubiquitously used by living cells, e.g., for cellular respiration, but also by immune cells for defense. Recent advances suggest that redox reactions can create a bridge between bio-molecular systems and electronic devices [79].

## VII. CONCLUSION

In this paper, the concept of photosynthetic molecules for the purpose of molecular communications in a nanoscale has been presented. Several of the unique properties of chlorophylls and carotenoids, that match very well with this application were highlighted: (a) effective energy harvesting of visible light in 380-700 nm region, (b) energy efficiency in signals transmission with energy per bit of about $2.0 \times 10^{-18}$ J, (c) transmission delays being on average only 20 ps, and (d) ability of signal conversion from photons to FRET and then to electrons. The results of two supporting experiments focused on carotenoid→chlorophyll signal transfer and the system energy efficiency were shown. In the former, based on the absorbance and fluorescence spectra, the FRET efficiency between carotenoids and chlorophylls, was determined. In the latter, on the basis of the fluorescence kinetic curves, the photochemical efficiency of the chlorophyll was calculated. This corresponds to the efficiency of converting FRET signals and absorbed photons into electrons during first stages of photosynthesis. Finally, the communication parameters of such a system estimating the respective bit error rate, energy consumption and discussing the possible throughput were analyzed. These parameters, especially energetic characteristics in terms of energy harvesting abilities and low consumption indicate a high potential of photosynthetic systems for future nanotechnology applications.

## APPENDIX: PREPARATION OF PLANT MATERIAL

Thylakoids were isolated from fresh leaves of spinach (Spinacia oleracea L.) according to the method described in [47] with minor modifications. Briefly, after removing stems from the spinach, leaves were homogenized in wash medium (pH=7.8) containing 400 mM sucrose, 50 mM TRIS-HCl, 10 mM NaCl and 5 mM $MgCl_2$. The ground spinach leaves were filtered and remaining suspension was centrifuged at 4200×g for 15 minutes in 6 $^0$C. After the supernatant was removed, the pellet was re-suspended with a buffer (pH=6.5) containing 20 mM Hepes, 5 mM $MgCl_2$ and 15 mM NaCl. This step was triplicated. Then thylakoids were suspended in a buffer containing 400 mM sucrose, 20 mM Hepes, 5 mM $MgCl_2$ and 15 mM NaCl to the total concentration of chlorophyll of 1 mg/ml. The concentration of chlorophyll was determined spectrophotometrically: 0.03 ml of the sample was suspended in 2.7 ml of 80% acetone and 0.27 ml of distilled water, centrifuged (2200 x g for 5 minutes) and then it was gently pipetted out. Its absorbance was measured at wavelengths that correspond to maximal absorption of chlorophyll-a (663 nm) and chlorophyll-b (645 nm). Thylakoids enriched in PSII complexes were solubilized on ice in the presence of detergent (15% Triton X-100) for 25 minutes in darkness. After that, they were centrifuged (62000 × g, 10 min, 4 $^0$C) 4-5 times until the transparent supernatant was obtained. The final chlorophyll concentration was adjusted to the total concentration of chlorophyll of 1.5 mg/ml. Only fresh samples (not frozen) were used for further experiments.

## ACKNOWLEDGMENT
The authors would like to thank Prof. K. Burda for her help at the initial stage of the project and Prof. L. Janowski for his suggestions regarding probability distributions.

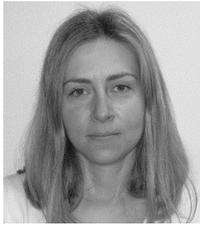
**ALEKSANDRA ORZECHOWSKA** received Ph. D. in biophysics from the Jagiellonian University in 2009. She is currently working as an assistant professor at the AGH University of Science and Technology in Krakow, Poland. She spent about 1 year in total as a post-doc at the Max Planck Institute of Molecular Plant Physiology, Australian National University, Commonwealth Scientific & Industrial Research Organisation (CSIRO) in Canberra and Research Center of Photon System Instruments in Czech Republic. Her main research interests are electron transport in photosynthetic reaction centers of type Q including plants and purple bacteria, molecular mechanisms induced by light in photosynthesis and abiotic stresses affected by plants. She was involved in several research projects, especially IATS, within the European FP7-SME Programme.

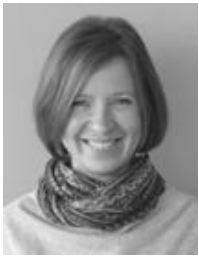
**JOANNA FIEDOR** received Ph.D. degree in biochemistry from the Jagiellonian University, Krakow, Poland in 2007. From 1997-1999 she stayed in Munich working at the Ludwig-Maximilians University (LMU), Germany, and in 2002 at the Kwansei Gakuin University, Sanda, Japan. Currently, she is an assistant professor at the AGH University of Science and Technology, Krakow, Poland. Her research interests are focused on natural biocompounds in relation to human health.

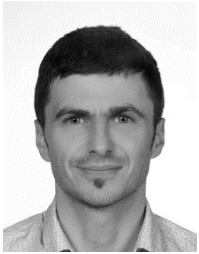
**PAWEL KULAKOWSKI** received Ph.D. in telecommunications from the AGH University of Science and Technology in Krakow, Poland, in 2007, and currently he is working there as an assistant professor. He spent about 2 years in total as a post-doc or a visiting professor at Technical University of Cartagena, University of Girona, University of Castilla-La Mancha and University of Seville. He was involved in research projects, especially European COST Actions: COST2100, IC1004 and CA15104 IRACON, focusing on topics of wireless sensor networks, indoor localization and wireless communications in general. His current research interests include molecular communications and nano-networks. He was recognized with several scientific distinctions, including 3 awards for his conference papers and a governmental scholarship for young outstanding researchers.